\documentclass[12pt]{article}

\usepackage{latexsym}
\usepackage{epsf}

\newcommand{\be}{\begin{equation}}
\newcommand{\ee}{\end{equation}}
\def\tr{{\rm tr}}
\newcommand{\Tr}{{\rm Tr}}

\def\tfrac#1#2{{\textstyle {#1 \over #2}}}
\def\dfrac#1#2{{\displaystyle {#1 \over #2}}}
\def\dint{\displaystyle \int }

\begin{document}
{

\flushright {\tt hep-th/0110203}

}

\vspace{5mm}
\begin{center}
{\bf\Large On Low-Energy Effective Action of Noncommutative
Hypermultiplet Model
}\\[1cm]

{\large\bf I.B. Samsonov\footnote{E-mail: samsonov@phys.tsu.ru}}\\[0.5cm]
{\it
  Department of Quantum Field Theory, Tomsk State University,
          Tomsk 634050, Russia}
\end{center}
\setcounter{footnote}{0}

\vspace{5mm}
\begin{center}
\bf Abstract
\end{center}
\begin{quotation}
\small
We consider the noncommutative hypermultiplet model within harmonic
superspace approach. The 1-loop four-point contributions to the effective
action of selfinteracting $q$-hypermultiplet are computed. This model
has two coupling constants instead of a single one in commutative case. It
is shown that both these coupling constants are generated by 1-loop
quantum corrections in the model of $q$-hypermultiplet interacting with
vector multiplet. The holomorphic effective action of
$q$-hypermultiplet in external gauge superfield is calculated.
For the fundamental representation there is no UV/IR-mixing and
the holomorphic potential is a $\star$-product generalization of a
standard commutative one. For the adjoint representation of $U(N)$ gauge
group the leading contributions to the holomorphic effective action
are given by the terms respecting for the UV/IR-mixing which are related to
$U(1)$ phase of $U(N)$ group.
\end{quotation}

\vspace{4mm}
Noncommutative field theory attracts much attention due to its remarkable
properties and profound links with modern string/brane activity. This
theory is based on an idea of underlying noncommutative space-time
geometry \cite{Connes,Seib99} that allows to incorporate a nonlocal
structure of the theory and a possibility to apply the standard quantum
field theory methods (see ref. \cite{Nekr} for review).

Noncommutative field theories possess specific properties of UV- and
IR-divergences (the UV/IR-mixing \cite{Min}) which distinguish them
from ordinary theories very essentially. For example, the
insertion of noncommutativity can break the renormalizability of field models
\cite{Arefeva}. 
The presence of nonplanar diagrams in gauge theories \cite{Zanon,Liu,Mehen}
leads to the deformation of $\star$-product by quantum corrections 
($\star_n$-products appear) that breaks the manifest gauge symmetry of
effective action. The restoration of manifest gauge invariance of effective
action involves the open Wilson line operators \cite{Liu,Mehen} which 
absorb all $\star_n$-products into gauge invariant objects.

Most of these problems are convenient to explore in the frame of
supersymmetric theories. The gauge invariance of effective action had been
examined by many authors on the example of noncommutative ${\cal N}=4$
SYM model \cite{Zanon,Liu} since this model is finite and have no
UV/IR-mixing \cite{Hash}. The structure of low-energy effective action in
the model of ${\cal N}=2$ SYM was studied in the papers
\cite{Arm,Armoni,Holl,Zanon,Sams1}, effective potential in noncommutative
Wess-Zumino model was considered in \cite{Petrov}.

Our aim is to investigate the low-energy effective action of ${\cal N}=2$
supersymmetric model of matter fields (hypermultiplet model) on
noncommutative plain. A natural approach for studying the ${\cal N}=2$
supersymmetric models is a harmonic superspace method \cite{harm}. It
provides an explicit ${\cal N}=2$ supersymmetry at all steps
of quantum computations. 
The ${\cal N}=2$ harmonic superspace is obtained from the conventional 
${\cal N}=2$ superspace by adding the harmonic variables which 
parameterize the $SU(2)/U(1)$ coset of internal automorphisms group of 
${\cal N}=2$ supersymmetry algebra (see \cite{Ivanov} for details).
The generalization of this construction to the case
of noncommutative geometry is straightforward. We suppose that only the
bosonic coordinates of superspace obey the Moyal-Weyl commutation
relations
\be
[x_\mu,x_\nu]=i\theta_{\mu\nu},
\label{e1}
\ee
where $\theta_{\mu\nu}$ is the constant antisymmetric tensor while the
fermionic coordinates of superspace are still anticommuting.
Such a deformation does not break supersymmetry since the derivatives 
$\partial_\mu$ are supercovariant. This is just the simplest way to introduce
the noncommutativity without breaking of supersymmetry. Note however some
attempts to construct a superspace with non-anticommuting Grassman 
coordinates \cite{Klemm} where the algebraic aspects of such deformations
were studied. A general case of deformations of superspace was considered 
also in \cite{Ferrara}. The harmonic variables
remain commutative because they correspond to the $SU(2)$
automorphisms algebra which in not deformed since 
the deformations in the bosonic sector only do not modify the supersymmetry
algebra \cite{Ferrara,Klemm}. The noncommutative ${\cal N}=2$ harmonic
superspace was considered firstly in \cite{Belh}.

A natural multiplication of superfields on noncommutative plain (\ref{e1}) 
is given by the $\star$-product
\be
(\phi\star\psi)(x)=\phi(x)e^{\frac i2
\theta^{\mu\nu}\overleftarrow{\partial_\mu}\overrightarrow{\partial_\nu}}
\psi(x),
\label{e2}
\ee
which is associative but noncommutative. 

In this letter we consider the noncommutative models of selfinteracting
$q$-hypermultiplet and $q$-hypermultiplet interacting with vector
superfield. In commutative case these theories were studied in
\cite{harm}-\cite{Sams}. We are interested mainly in new features of these
theories appearing due to the noncommutativity. The model of selfinteracting
commutative
$q$-hypermultiplet is known to be nonrenormalizable \cite{harm}, but it
was shown that the selfinteraction can be induced by quantum corrections
from the model of $q$-hypermultiplet interacting with vector multiplet
\cite{IKZ}. For the model of $q$-hypermultiplet in external gauge
superfield the low-energy effective action is defined by the holomorphic
potential which was computed exactly in papers \cite{Buch1_,Buch2} using the
harmonic superspace technique. In this paper we show how these results
can be generalized to noncommutative case.

To illustrate the features of noncommutative quantum field theory in
harmonic superspace we consider the simplest $q$-hypermultiplet model
with the quartic selfinteraction of the type 
$\lambda(\breve q^+)^2(q^+)^2$ \cite{harm}.
The action of corresponding noncommutative model is obtained by insertion of
$\star$-product (\ref{e2}) instead of usual multiplication
\footnote{
We follow the notations for harmonic superspace objects which
were accepted throughout the papers \cite{Kuz97}-\cite{Sams}.
}
\begin{eqnarray}
S[q,\breve q]&=&S_0[q,\breve q]+S_{int}[q,\breve q],
\label{e3a}
\\
 S_0[q,\breve q]&=&\dint d\zeta^{(-4)}\breve q^+D^{++}q^+,
\label{e3b} \\
S_{int}[q,\breve q]&=&
 \dint d\zeta^{(-4)}(\alpha\breve q^+\star\breve q^+\star q^+\star
q^++\beta \breve q^+\star q^+\star\breve q^+\star q^+).
\label{e3c}
\end{eqnarray}
Note that there is a possibility to introduce the two coupling constants
$\alpha,\beta$ instead of a single one $\lambda$ in commutative case due
to two types of ordering of superfields. This situation is similar to 
noncommutative $\phi^4$ model \cite{Arefeva}.
In commutative limit $\theta\rightarrow0$ the
two terms in the interaction (\ref{e3c}) reduce to a single one
$\lambda(q^+)^2(\breve q^+)^2$.

The propagator corresponding to the free action (\ref{e3b})
is given by \cite{harm}
\be
<\breve q(\zeta_1)q(\zeta_2)>=G_0^{(1,1)}(1|2)=-\frac1{\Box_1}\frac{(D^+_1)^4(D^+_2)^4\delta^{12}
(z_1-z_2)}{(u^+_1u^+_2)^3}.
\label{e4}
\ee
The action $S_{int}$ (\ref{e3c}) defines the four-point vertex (in momentum
space)
\be
(2\pi)^4\delta^4(p_1+p_2+p_3+p_4)
[\alpha \cos\frac{p_1\theta p_2}2
\cos\frac{p_3\theta p_4}2+\beta\cos(\frac{p_1\theta p_3}2+
\frac{p_2\theta p_4}2 )].
\label{e4.1}
\ee

We consider also the model of massive $q$-hypermultiplet with the
free action
\be
S_0[q,\breve q]=\dint d\zeta^{(-4)}\breve q^+(D^{++}+iV^{++}_0)q^+
\equiv\dint d\zeta^{(-4)}\breve q^+{\cal D}^{++}q^+,
\label{e5}
\ee
where $V_0^{++}=-\bar W_0(\theta^+)^2-W_0(\bar \theta^+)^2$,
$\bar W_0W_0=m^2$ is the mass. The corresponding propagator was found in
refs. \cite{Zupnik86,Zupnik87,IKZ} in the form
\be
<\breve q(\zeta_1)q(\zeta_2)>=G^{(1,1)}(1|2)=
 -\dfrac1{\Box_1+m^2}(D^+_1)^4(D^+_2)^4
\left\{
\frac{e^{i\Omega_0(1)-i\Omega_0(2)}\delta^{12}(z_1-
z_2)}{(u^+_1u^+_2)^3}
\right\},
\label{e6}
\ee
where $\Omega_0=-\bar W_0\theta^+\theta^--W_0\bar\theta^+\bar\theta^-$ is
a bridge superfield \cite{IKZ}.

An interaction of $q$-hypermultiplet (\ref{e5}) with the vector multiplet
is introduced by minimal way ${\cal D}^{++}\rightarrow{\cal D}^{++}+
iV^{++}\star$. For the fundamental representation of $U(1)$
gauge group it reads
\be
S_{int,f}=\dint d\zeta^{(-4)}\breve q^+\star iV^{++}\star q^+.
\label{e7}
\ee
The corresponding vertex is given by
\be
(2\pi)^4\delta^4(p+k_1+k_2)e^{-\frac i2k_1\theta k_2}.
\label{e8}
\ee
The $q$-hypermultiplet model (\ref{e5}) with the interaction (\ref{e7}) is
invariant under the following gauge transformations
\be
\begin{array}l
\delta q^+=i\lambda\star q^+,\qquad 
 \delta\breve q^+=-i\breve q^+\star\lambda,\\
\delta V^{++}=-D^{++}\lambda+i\lambda\star V^{++}-iV^{++}\star\lambda.
\end{array}
\label{e8_}
\ee
A generalization to the case of non-Abelian gauge group is trivial: the
vector superfield $V^{++}$ should be a matrix of $U(N)$ group while the
hypermultiplets $q^+$, $\breve q^+$ belong to the fundamental
representation. 

It is interesting to study also the adjoint representation
of gauge group when the interaction is given by
\be
S_{int,ad}=i\tr\dint d\zeta^{(-4)}\breve Q^+\star[V^{++},Q^+]_\star,
\label{e9}
\ee
where $[V^{++},Q^+]_\star=V^{++}\star Q^+-Q^+\star V^{++}$ and the
superfields $V^{++}$, $Q^+$, $\breve Q^+$ are the matrices of $U(N)$
group. For the case of adjoint representation the gauge transformations
look like
\be
\begin{array}l
\delta Q^+=i[\lambda,Q^+]_\star,\qquad 
 \delta\breve Q^+=-i[\breve Q^+,\lambda]_\star,\\
\delta V^{++}=-D^{++}\lambda+i[\lambda,V^{++}]_\star.
\end{array}
\label{e9_}
\ee

Now we begin with the quantum computations of 1-loop four-point diagram in
the model (\ref{e3a})
\be
\begin{array}c
\Gamma^{(1)}_4[\breve q^+,q^+]=\Gamma_s[\breve q^+,q^+]+
 \Gamma_t[\breve q^+,q^+],\\
\epsfxsize=2.5cm
\begin{array}l
\Gamma_s[\breve q^+,q^+]=\\
\vspace{0.8cm}\phantom{A}
\end{array}
 \epsfbox{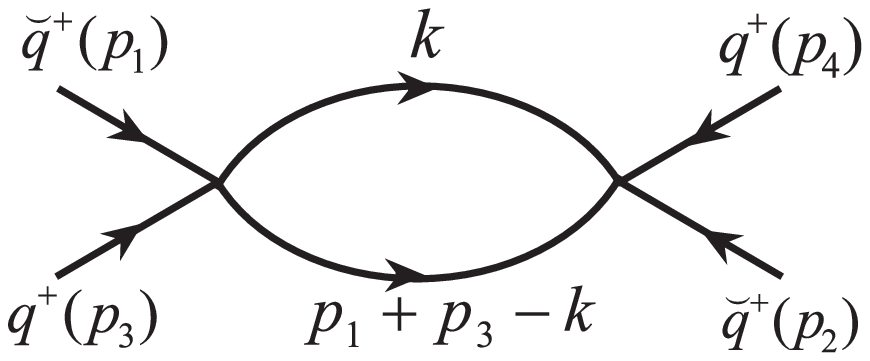},\qquad
\epsfxsize=2.5cm
\begin{array}l
\Gamma_t[\breve q^+,q^+]=\\
\vspace{0.8cm}\phantom{A}
\end{array}
\epsfbox{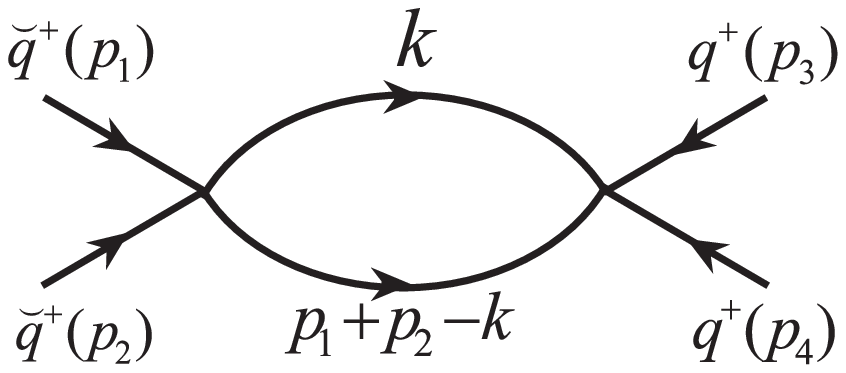}.
\end{array}
\label{e26}
\ee

\vspace{-0.8cm}\noindent
Using the Feynman rules (\ref{e4},\ref{e4.1}) we obtain the following
expressions for these diagrams
\be
\begin{array}{l}
\Gamma_{s,t}=\dint\tfrac{d^4p_1\ldots d^4p_4}{(2\pi)^{16}}
 d^4\theta^+_1 d^4\theta^+_2
 du_1du_2 \delta^4(p_1+p_2+p_3+p_4)\breve q^+(1)\breve
 q^+(2) q^+(3) q^+(4)\\
\ \times\tfrac{(D^+_1)^4(D^+_2)^4}{(u^+_1u^+_2)^3}\delta^8(\theta_1-\theta_2)
\tfrac{(D^+_1)^4(D^+_2)^4}{(u^+_1u^+_2)^3}
\delta^8(\theta_1-\theta_2)
I_{s,t}(p_1,\ldots,p_4),
\end{array}
\label{e27}
\ee
where
\be
\begin{array}l
I_{t}(p_1,\ldots,p_4)=
2\dint d^4k\frac{F_{t}(p_1,\ldots,p_4,k)}{k^2(p_1+p_2-k)^2},\\
I_{s}(p_1,\ldots,p_4)=
8\dint d^4k\frac{F_{s}(p_1,\ldots,p_4,k)}{k^2(p_1+p_3-k)^2},
\end{array}
\label{e28}
\ee
$F_{s,t}(p_1,\ldots,p_4,k)$ are the functions the structure of which is
stipulated by noncommutativity
\be
\begin{array}{rl}
F_t(p_1,\ldots,p_4,k)=&(\alpha\cos\frac{p_1\theta p_2}2
 \cos\frac{k\theta(p_1+p_2)}2+
 \beta\cos(\frac{p_1\theta k}2+\frac{p_2\theta(p_1-k)}2))\\
&\times(\alpha\cos\frac{k\theta(p_1+p_2)}2\cos\frac{p_3\theta p_4}2+
 \beta\cos(\frac{k\theta p_3}2-\frac{(p_3+k)\theta p_4}2)),\\
F_s(p_1,\ldots,p_4,k)=&(\alpha\cos\frac{k\theta p_1}2
 \cos\frac{(p_1-k)\theta p_3}2+\beta\cos(\frac{k\theta(p_1+p_3)}2+
  \frac{p_1\theta p_3}2))\\
&\times(\alpha\cos\frac{k\theta p_4}2\cos\frac{(p_4+k)\theta p_2}2+
 \beta\cos(\frac{p_2\theta p_4}2-\frac{k\theta(p_1+p_3)}2)).
\end{array}
\label{e29}
\ee
Further computations are very similar to usual ones in commutative
$q^4$-model given in \cite{harm}. The resulting expression for the
effective action (\ref{e27}) reads
\be
\Gamma_{s,t}=\dint\frac{d^4p_1\ldots d^4p_4}{(2\pi)^{16}}d^8\theta
du_1du_2
\dfrac{\breve q^+(1)\breve q^+(2) q^+(3)q^+(4)
}{(u^+_1u^+_2)^2}
\delta^4(p_1+p_2+p_3+p_4)I_{s,t}(p_1,\ldots,p_4).
\label{e30}
\ee
This effective action differs from the corresponding commutative one by the
presence of two functions $F_{s,t}$ in the integrals (\ref{e28}). These
functions change the UV-structure of integrals $I_{s,t}$ and lead to the
appearance of UV/IR-mixing.

To study the structure of momentum integrals (\ref{e28}) let us split the
functions $F_{s,t}$ into planar ({\it pl}) and non-planar ({\it npl}) parts
\be
F_{s,t}(p_1,\ldots,p_4,k)=F_{s,t}^{pl}(p_1,\ldots,p_4)
 +F_{s,t}^{npl}(p_1,\ldots,p_4,k),
\label{e31}
\ee
where we assume that $F_{s,t}^{pl}$ do not contain the terms depending
on the momentum $k$. The planar parts are written explicitly
\be
\begin{array}{l}
F_t^{pl}=\frac{\alpha^2}2\cos\frac{p_1\theta p_2}
 2\cos\frac{p_3\theta p_4}2,\\
F_s^{pl}=(\frac{\alpha^2}8+\frac{\beta^2}2)\cos(\frac{p_1\theta p_3}2+
 \frac{p_2\theta p_4}2)+\frac{\alpha\beta}2\cos(-\frac{p_1\theta p_3}2+
 \frac{p_2\theta p_4}2).
\end{array}
\label{e32}
\ee
Substituting the expressions (\ref{e32}) into integrals (\ref{e28})
\be
\begin{array}{l}
I^{pl}_{t}(p_1,\ldots,p_4)=
2F^{pl}_{t}(p_1,\ldots,p_4)\dint\frac{d^4k}{k^2(p_1+p_2-k)^2},\\
I^{pl}_{s}(p_1,\ldots,p_4)=8F^{pl}_{s}(p_1,\ldots,p_4)
\dint\frac{d^4k}{k^2(p_1+p_3-k)^2},
\end{array}
\label{e33}
\ee
one can see that the planar contributions have the same divergences as
corresponding diagrams of commutative $q$-hypermultiplet. The momentum
integrals in eq. (\ref{e33}) have the IR-divergences at low external momenta
which do not appear in massive theory (these singularities are not related
to the UV/IR-mixing) and UV-divergences which can not be renormalized
since the coupling constants $\alpha,\beta$
 have the mass dimension $-2$ as in commutative
theory \cite{harm}. So, the selfinteracting noncommutative
$q$-hypermultiplet model is nonrenormalizable as well as the
corresponding commutative theory. Inserting the noncommutativity does not
improve the situation with renormalizability here.

Let us study now the structure of non-planar diagrams where the effect of
UV/IR-mixing arises. One can show
that all non-planar terms of functions (\ref{e29}) look like
$$
\cos\frac{p_1\theta p_2}2\cos\frac{p_3\theta p_4}2\cos(k\theta(p_1+p_2))
$$
with various combinations of external momenta. They define the structure
of momentum integrals of non-planar type
\be
\begin{array}{l}
I^{npl}_t(p_1,\ldots,p_4)\sim
 \cos\frac{p_1\theta p_2}2\cos\frac{p_3\theta p_4}2
 \dint d^4k\frac{e^{ik\theta(p_1+p_2)}}{k^2(p_1+p_2-k)^2}\\
=\cos\frac{p_1\theta p_2}2\cos\frac{p_3\theta p_4}2
\dint_0^1d\xi
K_0(\sqrt{\xi(1-\xi)P}),
\end{array}
\label{e34}
\ee
where
$
P=(p_1+p_2)^2\cdot(p_1+p_2)\circ(p_1+p_2),
$
$p_1\circ p_2=p_1^\mu(\theta\theta)_{\mu\nu}p_2^\nu$, $K_0$ is the
modified Bessel function. This expression has no UV-divergence but it is
singular at low momenta $p_i\rightarrow0$ due to the asymptotics of Bessel
function
\be
K_0(x)\stackrel{x\rightarrow0}{\longrightarrow}-\ln\frac x2+{\rm finite}.
\label{e35}
\ee
Therefore at low external momenta the integral over momentum $k$ behaves as
\be
\int d^4k\frac{e^{ik\theta(p_1+p_2)}}{k^2(p_1+p_2-k)^2}\sim
\ln\frac{1}{P}.
\label{e36}
\ee
The expression (\ref{e36}) is singular in commutative limit
$\theta\rightarrow0$. Such a singularity of effective action was called
the UV/IR-mixing \cite{Min}. Therefore the effective action of 
noncommutative hypermultiplet does not reduce to a standard one in 
commutative limit.

As a result, the model of noncommutative hypermultiplet is
nonrenormalizable and has the UV/IR-mixing in the sector of non-planar
diagrams.

To obtain a renormalizable model in commutative case it was suggested
\cite{IKZ}
to consider the model (\ref{e5}) and generate the interaction like
(\ref{e3c}) by 1-loop low-energy corrections to the effective action of
the type
\be
\begin{array}c
\Gamma_4=\Gamma_s+\Gamma_t,\\[1mm]
\begin{array}r
\Gamma_s=\\
\vspace{8mm}\vphantom{A}
\end{array}
\epsfxsize=2.5cm
\epsfbox{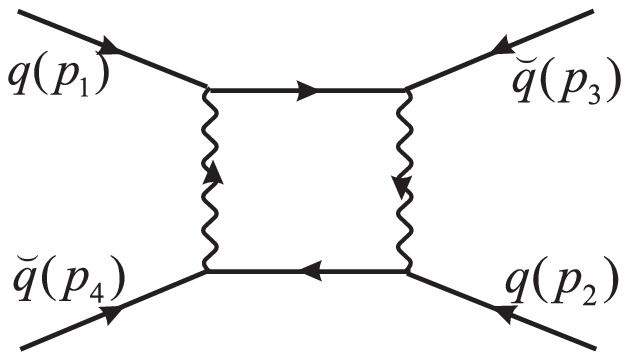},\qquad
\begin{array}r
\Gamma_t=
\vspace{1.3cm}\vphantom{A}
\end{array}
\epsfxsize=2.5cm
\epsfbox{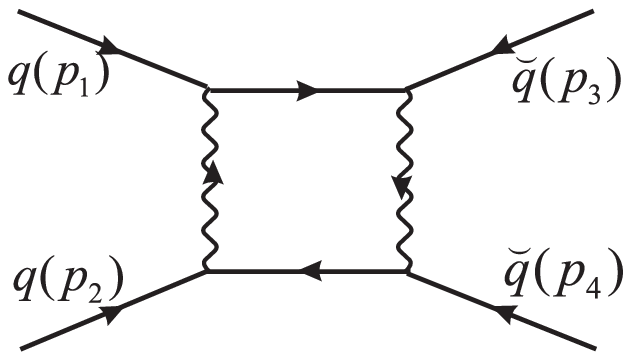}.\\[-0.2cm]
\end{array}
\label{e45}
\ee

\vspace{-0.5cm}\noindent
We study the analogical possibility in noncommutative hypermultiplet model.
In the low energy approximation the loops shrink down to points and 
we will see that the
finite parts of these diagrams give the coupling constants $\alpha,\beta$
of selfinteracting $q$-hypermultiplet (\ref{e3c}) induced only by quantum
corrections. Induced selfinteraction in the commutative hypermultiplet
model was considered in \cite{IKZ}.

To compute these diagrams we apply the Feynman rules obtained
(\ref{e6},\ref{e8}) and the following expression for the propagator of
intermediate gauge boson \cite{harm}
\be
i<V^{++}(1)V^{++}(2)>=\frac1{\vphantom{A}\Box_1}(D^+_1)^4
\delta^{12}(z_1-z_2)\delta^{(-2,2)}(u_1,u_2).
\label{VV}
\ee
Using the explicit expressions for the propagators and vertices one obtains
the following representations for the diagrams (\ref{e45})
\be
\begin{array}c
\Gamma_{s,t}[\breve q^+,q^+]=
 -\frac{ig^4}{(2\pi)^{16}}\dint d^4p_1d^4p_2d^4p_3d^4p_4
 F_{s,t}(p_1,\ldots,p_4)I_{s,t}(p_1,\ldots,p_4)\\\times\dint
 d^8\theta\frac{du_1du_2}{(u^+_1u^+_2)^2}q^+(p_1)q^+(p_2)
 \breve q^+(p_3)\breve q^+(p_4)
e^{i\Omega_0(2)-i\Omega_0(1)}e^{i\Omega_0(2)-i\Omega_0(1)},
\end{array}
\label{e46}
\ee
where
\be
\begin{array}l
I_s(p_1,\ldots,p_4)=\frac1{(2\pi)^4}
\dint \frac{d^4k}{k^2(k+p_1+p_3)^2}
\frac1{((k+p_1)^2+m^2)((k-p_4)^2+m^2)},
\\
I_t(p_1,\ldots,p_4)=\frac1{(2\pi)^2}
\dint \frac{d^4ke^{ik\theta(p_1+p_3)}}{
(k+p_3)^2(k-p_1)^2}
\frac1{(k^2+m^2)((k+p_3+p_4)^2+m^2)},
\end{array}
\label{e47}
\ee
\be
F_s(p_1,\ldots,p_4)=e^{-\frac i2p_1\theta p_3}e^{-\frac i2p_2\theta p_4},
\qquad
F_t(p_1,\ldots,p_4)=e^{-\frac i2p_2\theta p_1}e^{-\frac i2p_3\theta p_4}.
\label{e48}
\ee
It should be noted that the momentum integrals (\ref{e47}) are UV-finite.
Therefore there is no UV/IR-mixing and these diagrams are well defined in
the limits $\theta\rightarrow0$ and $p_i\rightarrow0$. At zero external
momenta both these integrals are computed exactly:
\be
I_{s,t}=\frac1{(4\pi)^2m^2}\left(
\frac1{m^2}\ln(1+\frac{m^2}{\Lambda^2_{s,t}})-\frac1{\Lambda^2_{s,t}+m^2}
\right),
\label{e49}
\ee
where $\Lambda_{s,t}$ are the parameters of infrared cutoff
which do not coincide generally speaking. The values of
$\Lambda_{s,t}$ should be taken less then the mass of the lightest particle 
that corresponds to the Wilsonian low-energy effective action.

In order to extract the relevant low-energy contribution from eq.
(\ref{e46}) it is necessary to employ the covariant derivatives algebra
\cite{harm}
\be
[{\cal D}^{++},{\cal D}^{--}]=D^0,
\label{e50}
\ee
where
\be
{\cal D}^{\pm\pm}=D^{\pm\pm}+iV_0^{\pm\pm},\qquad
 V_0^{\pm\pm}=D^{\pm\pm}\Omega_0.
\label{e51}
\ee
The relation (\ref{e50}) is used to prove the identity
\be
\begin{array}l
q^+\star q^+\star\breve q^+\star\breve q^+
 e^{i\Omega_0(2)-i\Omega_0(1)} e^{i\Omega_0(2)-i\Omega_0(1)}
 \\
\qquad=\frac12 [{\cal D}_1^{++},{\cal D}_1^{--}]
 q^+\star q^+\star\breve q^+\star\breve q^+
e^{i\Omega_0(2)-i\Omega_0(1)} e^{i\Omega_0(2)-i\Omega_0(1)}.
\end{array}
\label{e52}
\ee
Then it is easy to argue that only the first term
$\sim {\cal D}^{++}{\cal D}^{--}$ in this identity gives the leading
low-energy contribution. Integrating by parts one can cancel the harmonic
distribution $\frac1{(u^+_1u^+_2)^2}$ by the use of another identity
\cite{harm}
\be
D^{++}_1\frac1{(u^+_1u^+_2)^2}=
 D^{--}_1\delta^{(2,-2)}(u_1,u_2).
\label{e53}
\ee
The harmonic $\delta$-function on the right hand side of eq. (\ref{e53})
removes one of the two remaining harmonic integrals and allows us to
rewrite eq. (\ref{e46}) to the form
\be
\Gamma_{s,t}[\breve q^+,q^+]=-ig^4 I_{s,t}\cdot S_{s,t}[\breve q^+,q^+],
\label{e54}
\ee
where
\be
\begin{array}l
S_s=\dint d^4xd^8\theta du {\cal D}^{--}{\cal D}^{--}
  q^+\star\breve q^+\star q^+\star\breve q^+,\\
S_t=\dint d^4xd^8\theta du {\cal D}^{--}{\cal D}^{--}
  q^+\star q^+\star\breve q^+\star\breve q^+.
\end{array}
\label{e55}
\ee
The integrals over full ${\cal N}=2$ superspace in eqs. (\ref{e55}) can be
transformed to ones over the analytic subspace
\be
\begin{array}l
S_s=-2m^2\dint d\zeta^{(-4)} q^+\star\breve q^+\star q^+\star\breve q^+,\\
S_t=-2m^2\dint d\zeta^{(-4)} q^+\star q^+\star\breve q^+\star\breve q^+,
\end{array}
\label{e56}
\ee
where we have used the explicit form for $v^{--}$ given by eq. (\ref{e51})
and relation $\bar WW=m^2$.
As a result, the low-energy part of the effective action (\ref{e54})
is represented in the form of four-point $q$-hypermultiplet interaction
(\ref{e5})
\be
\Gamma_4=\int d\zeta^{(-4)}(\lambda_t
  \breve q^+\star\breve q^+\star q^+\star q^++
\lambda_s\breve q^+\star q^+\star\breve q^+\star q^+),
\label{e57}
\ee
where the coupling constants $\lambda_{s,t}$ appear due to quantum
corrections (\ref{e49}) as
\be
\lambda_{s,t}=\frac{g^2}{(4\pi)^2}\left(
\frac1{m^2}\ln(1+\frac{m^2}{\Lambda_{s,t}})-\frac1{\Lambda_{s,t}^2+m^2}
\right).
\label{e58}
\ee

Note that the effective action (\ref{e57}) is smooth in the limit
$\theta\rightarrow0$ and both induced coupling constants $\lambda_{s,t}$
(\ref{e58}) reduce to a single coupling constant obtained in ref. \cite{IKZ}
for the commutative hypermultiplet model.

Let us consider now the low-energy effective action of massive single
$q$-hypermultiplet model (\ref{e5}) in fundamental representation of
$U(1)$ gauge group. The 1-loop effective action in Coulomb branch
is given by
\be
\begin{array}c
\Gamma^{(1)}[V^{++}]=i\tr\ln(\nabla^{++}\star)=
 \sum\limits_{n=1}^\infty\Gamma_n,\\
\Gamma_n=i\frac{(-i)^n}{n}\dint d\zeta_1^{(-4)}\ldots d\zeta_n^{(-4)}
G^{(1,1)}(\zeta_1,\zeta_2)\star V^{++}(\zeta_2)\ldots
G^{(1,1)}(\zeta_n,\zeta_1)\star V^{++}(\zeta_1),
\end{array}
\label{e79}
\ee
where $\nabla^{++}\star=D^{++}+iV^{++}_0+iV^{++}\star$.
 In momentum space the $n$-point
function $\Gamma_n$ reads
\be
\begin{array}l
\Gamma_n=\frac{i(-i)^n}{n(2\pi)^{4n}}\dint d^4p_1
 d^4\theta^+_1du_1
\ldots d^4p_nd^4\theta^+_ndu_n
\dint d^4k\delta^4(\sum p_i)e^{-\frac i2k\theta(\sum p_i)}\\
\times G^{(1,1)}(k+p_2+p_3+\ldots+p_n)G^{(1,1)}(k+p_3+\ldots+p_n)\ldots
 G^{(1,1)}(k+p_n)G^{(1,1)}(k)\\
\times e^{-\frac i2\sum^n_{i<j}
 p_i\theta p_j}V^{++}(1)\ldots V^{++}(n).
\end{array}
\label{e80}
\ee
The first exponent in eq. (\ref{e80}) reduces to unity due to the relation
\be
e^{-\frac i2k\theta(\sum p_i)}\delta^4(\sum p_i)=\delta^4(\sum p_i).
\label{e80_}
\ee
The equation (\ref{e80_}) ensures the absence of nonplanar diagrams and
UV/IR-mixing. Therefore, the limit $\theta\rightarrow0$ is smooth and in
low-energy approximation (when we neglect all derivatives of $V^{++}$)
the factor $e^{-\frac i2\sum_{i<j}^np_i\theta p_j}$ can be dropped.
In such an approximation the effective action of noncommutative
$q$-hypermultiplet reduces to a standard one of commutative
$q$-hypermultiplet calculated in ref. \cite{Buch2}.
A non-trivial result here is the absence of non-planar diagrams
and UV/IR-mixing.

Now consider the next order in the approximation when all derivatives
related to noncommutativity are kept. In this case we keep the
factor $e^{-\frac i2\sum_{i<j}^np_i\theta p_j}$ in eq. (\ref{e80})
and use it to restore the $\star$-product of superfields due to the
identity
\be
\begin{array}c
\dint d^4p_1\ldots d^4p_n
e^{-\frac i2\sum_{i<j}^np_i\theta p_j}
 V_1(p_1)\ldots V^{++}_n(p_n) \delta^4(\sum p_i)\\=
\dint d^4xV_1^{++}(x)\star V_2^{++}(x)\star\ldots\star V^{++}_n(x).
\end{array}
\label{rel}
\ee
We expect the resulting expression for the effective action to be gauge
invariant.
It means that we have to express the effective action
in terms of strength superfield which is a
nonlinear function of $V^{++}$ even for the Abelian gauge group (it can be found in
complete analogy with commutative case \cite{Zupnik87})
\be
W=\frac14\bar D^+_{\dot\alpha}\bar D^{+\dot\alpha}
 \sum\limits^\infty_{n=1}i(-ig)^{n}\int du_1\ldots du_n
\frac{V^{++}(z,u_1)\star V^{++}(z,u_2)\star\ldots\star
 V^{++}(z,u_n)}{(u^+u^+_1)(u^+_1u^+_2)\ldots(u^+_nu^+)}.
\label{e9.4}
\ee
To
simplify this problem we choose the external gauge superfield in
the form
\be
\tilde W=-\frac14\int du\bar D^-_{\dot\alpha}
 \bar D^{-\dot\alpha}\tilde V^{++}(z,u),
\label{e82}
\ee
that corresponds to the first term in series (\ref{e9.4}).
With such a special background $\tilde W$ one obtains the standard expression
for holomorphic effective action \cite{Buch2}
written in terms of strength superfields $\tilde W$ where conventional
multiplication should be replaced by $\star$-product due to the relation
(\ref{rel}).
To return to arbitrary background it is necessary to restore the full
strength $W$ from $\tilde W$ with the help of gauge transformations
$$
W=e^{i\lambda}_\star\star\tilde W\star e^{-i\lambda}_\star
$$
with a special $\lambda=\lambda(W)$. Therefore, for arbitrary $W$
the holomorphic effective action of noncommutative
$q$-hypermultiplet reads
\be
\Gamma[V^{++}]=-\frac1{64\pi^2}\int d^4xd^4\theta
W\star W\star\ln_\star\frac{W}{\Lambda}+c.c.
\label{e84}
\ee
Note that this expression is manifestly gauge invariant and has correct
commutative limit.

A generalization of this result to the case of $U(N)$ gauge group broken
down to $[U(1)]^N$ is rather trivial since the strength $W$ belonging to
$[U(1)]^N$ can be chosen as $W={\rm diag}(W_1,\ldots,W_N)$. Therefore the
effective action of such theory is a sum of actions (\ref{e84})
\be
\Gamma_{[U(1)]^N}[V^{++}]=\sum_{i=1}^N\Gamma_{U(1)}[V^{++}_i],
\label{e85}
\ee
where $\Gamma_{U(1)}[V^{++}_i]$ is given by eq. (\ref{e84}).

In the case of adjoint representation of gauge group the situation is more
complicated. Let us consider the classical action of $q$-hypermultiplet in
adjoint representation when the vector superfield belongs to
the Cartan subalgebra of $u(N)$, i.e.
\be
Q^+=\sum\limits^N_{i,j}q^+_{ij}e_{ij},
\qquad
\breve Q^+=\sum\limits^N_{i,j}\breve q^+_{ij}e_{ji},\qquad
V^{++}=\sum\limits^N_{k=1}V^{++}_ke_{kk},
\label{e86}
\ee
where
\be
(e_{ij})_{kl}=\delta_{ik}\delta_{jl},\qquad i,j,k,l=1,\ldots,n
\label{basis}
\ee
is a Cartan-Weyl basis of $u(N)$ algebra \cite{Barut}. One can easily check
that the interaction (\ref{e9}) now looks like
\be
\tr\dint d\zeta^{(-4)}\breve Q^+\star[V^{++},Q^+]_\star=
\sum\limits^N_{k,l}\dint d\zeta^{(-4)}
  (\breve q^+_{kl}\star V^{++}_k\star q^+_{kl}
 -\breve q^+_{kl}\star q^+_{kl}\star V^{++}_l).
\label{e87}
\ee
Therefore the contributions from hypermultiplets $q^+_{ij}$, $q^+_k$
factorize and the formal expression for the 1-loop effective action reads
\be
\Gamma[V^{++}]=\sum\limits^N_{k,l=1}\Tr\ln\nabla^{++}_{kl}.
\label{e88}
\ee
where the operators $\nabla^{++}_{kl}$, act on
hypermultiplets by the rule
$$
\nabla^{++}_{kl}q^+_{kl}={\cal D}^{++}q^+_{kl}+iV^{++}_k\star q^+_{kl}-
  iq^+_{kl}\star V^{++}_l.
$$
The effective action (\ref{e88}), unlike the fundamental representation,
contain the terms respecting for the UV/IR-mixing which appear in
two-point diagrams. These diagrams can be computed in complete analogy as
it was done in the case of fundamental representation. The resulting
expression for the nonplanar part of two-point one-loop effective action
reads
\be
\begin{array}c
\Gamma_2^{np}[V^{++}]=\dfrac1{16\pi^2}\dint d^4pd^8\theta
 du_1du_2\dint_0^1d\xi K_0(\sqrt{(\xi(1-\xi)p^2+m^2)p\circ p})\\
\times \dfrac{{\cal V}^{++}(p,\theta,u_1)
  {\cal V}^{++}(-p,\theta,u_2)}{(u^+_1u^+_2)^2},
\end{array}
\label{e89}
\ee
where ${\cal V}^{++}=\sum_{i=1}^NV^{++}_i$ is the superfield corresponding
to $U(1)$ subgroup of $U(N)$ gauge group.
In the low-energy approximation we
employ the asymptotics of the Bessel function (\ref{e35}) to extract
the part of the effective action which is singular in the commutative
limit $\theta\rightarrow0$
\be
\Gamma_2^{np}[V^{++}]=\frac1{32\pi^2}\int d^4pd^8\theta
 du_1du_2\ln\frac4{m^2p\circ p}\frac{{\cal V}^{++}(p,\theta,u_1)
  {\cal V}^{++}(-p,\theta,u_2)}{(u^+_1u^+_2)^2}.
\label{e90}
\ee
We see that the UV/IR-mixing appears only in the $U(1)$ sector that was
firstly observed for noncommutative $U(N)$ Yang-Mills model in \cite{Arm}
and for supersymmetric gauge theories in \cite{Armoni,Holl}.

The next order contribution to the effective action which is finite in
the limit $\theta\rightarrow0$ corresponds to the holomorphic potential
of commutative $q$-hypermultiplet in adjoint representation of $SU(N)$
group calculated in \cite{Sams}.

Let us summarize the results.
We have considered the model of noncommutative selfinteracting 
hypermultiplet using the harmonic superspace approach.
It is worth pointing out that such a model has two coupling constants 
instead of a single one in commutative case.
The 1-loop four-point contributions to the effective action
of selfinteracting $q$-hypermultiplet are calculated. The ultraviolet
divergences in this model are responsible for the appearance of UV/IR-mixing terms.
These divergences can not be renormalized as well as in commutative case.
We find that both these coupling constants
are generated by 1-loop quantum corrections in the
model of $q$-hypermultiplet interacting with vector multiplet.

We have calculated the low-energy effective action of noncommutative
$q$-hypermultiplet in external gauge superfield. For the fundamental
representation of $U(N)$ gauge group the effective action is found to be
free of nonplanar contributions and UV/IR-mixing. This allows to calculate
the holomorphic potential using the harmonic superspace
approach and to represent it in manifestly gauge invariant form. We find
that the holomorphic effective action is exactly the $\star$-product
generalization of corresponding commutative one.

For the case of adjoint representation of $U(N)$ group the situation is
different. It is shown that the leading terms in the low-energy effective
action are those which are responsible for the UV/IR-mixing in this model.
The UV/IR-mixing appears only in $U(1)$ sector of $U(N)$ group while the
$SU(N)$ part is responsible for the contributions to the holomorphic
potential. 

In conclusion let us briefly comment on the gauge invariance of the effective
action of $q$-hypermultiplet in adjoint representation. At first sight the
contribution (\ref{e90}) does not preserve the gauge symmetry since it
contains the $\ln\frac4{m^2p\circ p}$ factor which does not allow to restore
the $\star$-product of superfields in the effective action. It is the common
problem of noncommutative gauge theories that the effective action can not be
expressed in terms of $\star$-product only but it involves the generalized
$\star$-multiplications of fields ($\star_n$-products) \cite{Zanon,Liu,Mehen}.
It is clear that the effective action must be gauge invariant since we start
from the gauge invariant classical theory. In the papers \cite{Liu,Mehen}
it was shown
that the gauge invariant effective action is built in terms of open Wilson
line operators which absorb all the $\star_n$-products into gauge invariant
objects. Moreover, in the recent papers \cite{Raam,ArmNew}
it was proved that manifestly 
singular logarithmic factors, like in eq. (\ref{e90}), do not spoil the 
gauge invariance but they are included into gauge invariant constructions
with help of open Wilson line operators. We are faced with the same situation
of logarithmic divergences in noncommutative hypermultiplet model in adjoint
representation. But it is worth pointing out that this problem does not appear 
for the $q$-hypermultiplet in fundamental representation since this model is
free of nonplanar diagrams.

The author is very grateful to I.L. Buchbinder for the supervision
and careful reading of the paper. I would like also to thank
B.M. Zupnik for valuable discussions and A. Armoni for useful comments.
The work was supported in part by INTAS grant, INTAS-00-00254.


\end{document}